\input{aipcheck}

\documentclass[
    ,final            
  ]
  {aipproc}

\layoutstyle{8x11single}

\begin{document}

\title{Energy density fluctuations in Early Universe}


\author{G.L. Guardo}{
  address={Department of Physics and Astronomy, University of Catania, Catania, Italy}
}

\author{V. Greco}{
  address={Department of Physics and Astronomy, University of Catania, Catania, Italy}
  ,altaddress={INFN - Laboratori Nazionali del Sud, Catania, Italy} 
}

\author{M. Ruggieri}{
  address={Department of Physics and Astronomy, University of Catania, Catania, Italy}
}

\begin{abstract}
The primordial nucleosinthesys of the element can be influenced by the transitions of phase that take place after
the Big Bang, such as the QCD transition.
In order to study the effect of this phase transition, in this work we compute the time evolution of thermodynamical quantities  
of the early universe, focusing on temperature and energy density fluctuations,
by solving the relevant equations of motion using as input the lattice QCD equation of state
to describe the strongly interacting matter in the early universe plasma.  
We also study the effect of a primordial strong magnetic field
by means of a phenomenological equation of state.
Our results show that small inhomogeneities of strongly interacting matter 
in the early Universe are moderately damped during the crossover.
\end{abstract}

\maketitle


\section{Introduction}
The history of the early universe left an imprint on the currently observed cosmos. 
According to the Friedmann equations, the temperature $T$ of the early universe plasma 
decreased as $a^{-1}$ with $a$ the scale parameter,
so that several phase transitions occurred during the early evolution of the universe,
like the electroweak phase transition at t$\approx$10$^{-10}$ sec (corresponding to T$\approx$100 GeV) and 
the QCD phase transition at t$\approx$10$^{-5}$ sec (T$\approx$ 150 MeV), 
then passing through the Big Bang Nucleosynthesis (BBN) at t=1 sec-3 min (T =0.1-1 MeV). 
This BBN plays a crucial role in constraining our views of the universe: 
it is essentially the only probe for the radiation dominated
epoch during the range $\approx$1-10$^{4}$ sec.
Moreover, the aforementioned phase transitions could strongly affect the BBN \cite{yagi}.

In this work, we focus on the role of the QCD phase transition on the evolution of the young universe,
with particular reference to the effect of the former on the evolution of the energy density fluctuations.
We also show preliminary results obtained considering the effect of a primordial strong
magnetic background on the aforementioned fluctuations.  
The primordial QCD phase transition is interesting because at that time a big change 
of the number of degree of freedom took place. The role of this primordial transition
is still matter of debate. For example, in \cite{applegate} it was realized 
that an inhomogeneous distribution of baryons due to a first-order QCD transition 
might change the primordial abundances of the light elements \cite{kajino}.

\section{Equations of state for strongly interacting matter}
In this work we describe the plasma in the early universe as a quark-gluon plasma plus electroweak matter
in thermal equilibrium. The energy density (as well as other thermodynamical quantities) are then written as
\begin{equation}
\varepsilon = \varepsilon_{ew} + \varepsilon_{qgp} ,
\label{eq:e1}
\end{equation}
where $\varepsilon_{ew}$, $\varepsilon_{qgp}$ correspond to the energy densities of the electroweak and
QCD plasmas respectively. The electroweak sector will be considered as a perfect gas of massless
particles in thermal equilibrium at the temperature $T$, hence
\begin{equation}
	\varepsilon_{ew}=g_{ew}\frac{\pi^{2}}{30}T^{4},\,\,\,\,\,\,\,
  P_{ew}=g_{ew}\frac{\pi^{2}}{90}T^{4}~,
\end{equation}
with $g_{ew}=$14.45 \cite{yagi}.
On the other hand, for the QCD plasma we cannot ignore the strong interactions which in particular
are very important in proximity of the phase transition. Therefore we need to use the appropriate equation of state
for this matter.

The simplest way to take into account the properties of the QGP is the EoS of the MIT bag Model \cite{bag}. 
In the bag model one describes the short-distance dynamics by a collisionless gas of massless quarks and gluons,
while the long-distance confinement effects are taken into account by subtracting a positive contribution to the pressure 
named the bag constant $B$. We present one result obtained within the bag model in this Section,
for a standard value of the Bag constant $B$=235 MeV$^{4}$, which is shown only for a matter of comparison with the
more realistic cases discussed below.

As an example of realistic EoS for the QCD plasma we use the one recently computed on the lattice 
for the case of 2+1 flavor QCD \cite{lattice}. 
Moreover, during the QCD transition strong magnetic fields
might have been present \cite{Campanelli:2013mea,Vachaspati:1991nm}. 
In order to take into account this possibility we present our preliminary results
for the energy density fluctuations in presence of a strong magnetic background,
using an effective model to compute the relevant equation of state \cite{Gatto:2010pt}. 
Here we consider an admittedly large value of the magnetic field,
$B\approx 10^{15}$ Tesla; this value of $B$ is chosen to estimate the maximum effect
a magnetic field can have on the evolution of the energy density fluctuations in the early universe.
The use of an effective model to describe the QCD plasma in presence of a magnetic background might 
appear opinable, since such model fails to reproduce the dependence of the
chiral critical temperature on the magnetic field strength \cite{Bali:2011qj}
(see however \cite{Chao:2013qpa} and also the discussions in \cite{Kojo:2012js}).
However our main point is to discuss the effect of the strengthening
of the phase transition with the strength of $B$, a characteristic which is observed
both in lattice simulations and in the model calculation, which is not related
to the dependence of the critical temperature on $B$. 

\begin{figure}
\includegraphics[width=7cm]{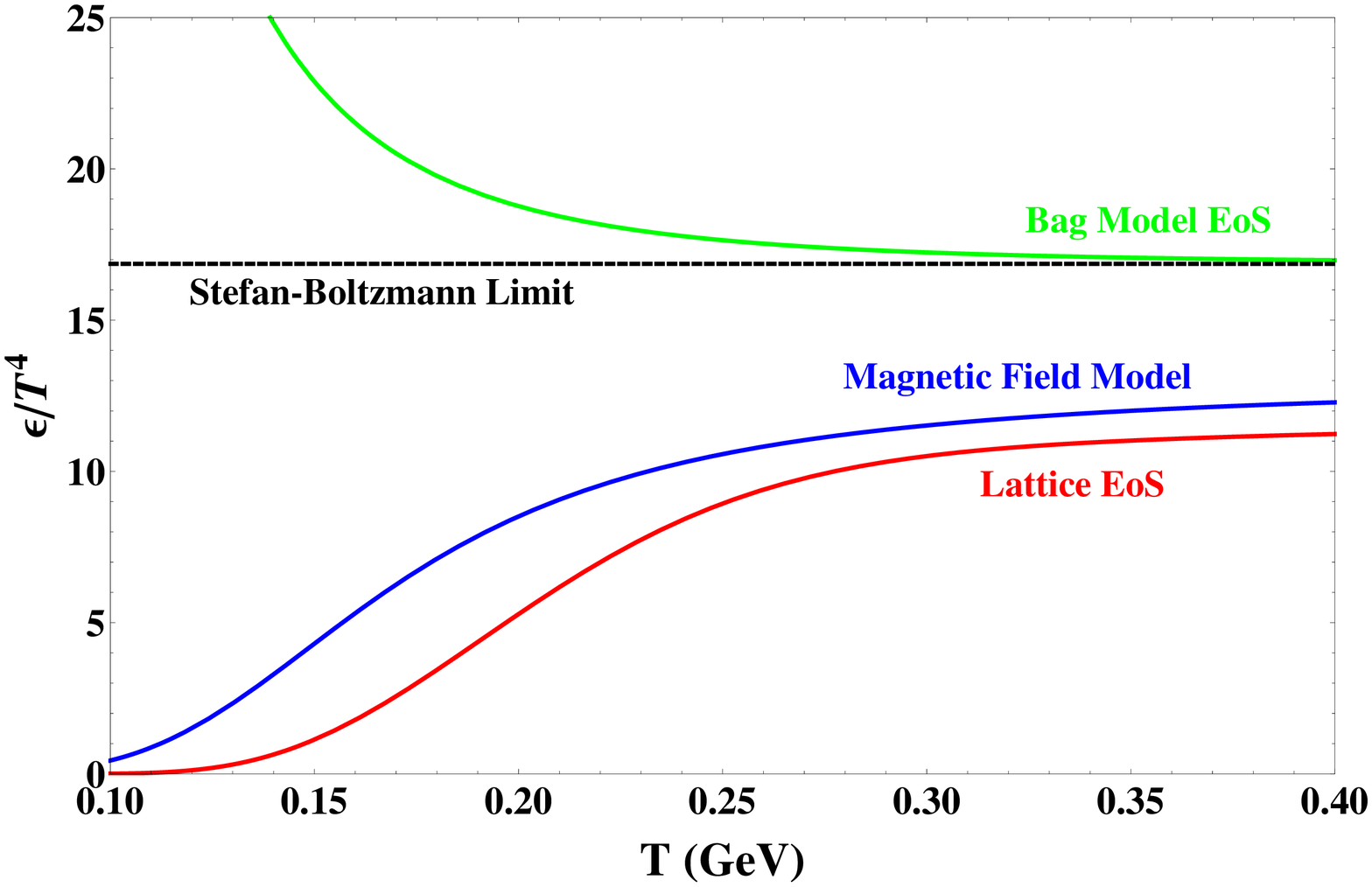}~~~\includegraphics[width=7cm]{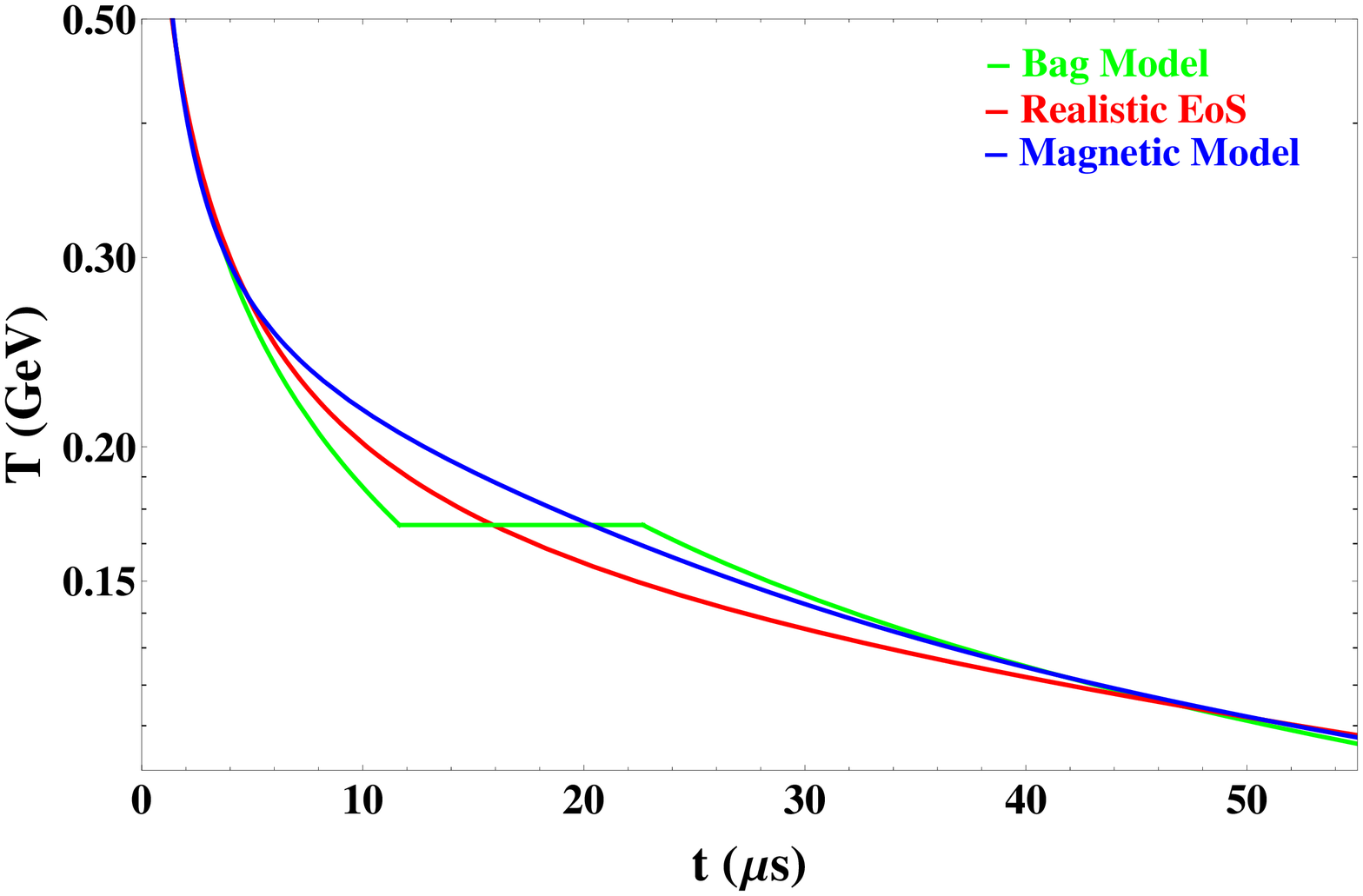}
\caption{{Left panel}. Energy density of the QCD plasma against temperature. 
Red line represents the lattice EoS, the blue line the same realistic EoS with the inclusion 
of a strong primordial magnetic field, and the green line corresponds to the bag model EoS. 
For reference the Stefan-Boltzmann limit is reported with the black dashed line.
{Right panel}. Temperature as a function of time for the three EoS discussed in the
main text. Color convention is the same of the left panel.}
\label{eos}
\end{figure}

In Fig. \ref{eos} we plot $\varepsilon_{qgp}$ as a function of temperature. 
The black dashed line corresponds to the case of ideal gas of massless particles 
(the Stefan-Boltzmann limit) that is directly correlated to the number of degrees of freedom of the system. 
Red solid line corresponds to the case of the lattice QCD EoS; blue line to the EoS with magnetic field.
For a matter of comparison we also show the case of the bag model by the solid green line.
We remind that even in the most recent textbooks the temperature evolution of the Early Universe is usually discussed with a simple Bag model.
The difference between the realistic EoS and the Stefan-Boltzmann limit indicates that the 
interactions in the plasma are non-negligible at temperatures which are relevant in our study temperatures.

\section{Temperature evolution}
In this Section we show our results for the temperature evolution in the early universe.
To determine this quantity we 
solve the Friedmann equation that, assuming that the expansion of the Universe is isentropic, takes the form
\begin{equation}
	\frac{d\varepsilon}{dt}=-3\sqrt{\frac{8\pi G\varepsilon}{3}}(\varepsilon+P)
\end{equation}
The numerical solutions for the equations of state described in the previous Section
are shown in the right panel of Fig. \ref{eos} that clearly shows how changing the equation of state
affects the behaviour of temperature against time. 
The Bag Model (green line) presents a plateau during the transition corresponding to the critical temperature of about 170 MeV,
at which the phase transition is of first order.
On the other hand because the QCD phase transition is actually a crossover, the use of the lattice EoS (red line) 
results in a smooth change of the temperature as a function of time. 
The magnetic field does not change the nature of the phase transition (blue line). 
We stress that the difference between the three cases are strictly confined to the time interval 
in which the transition takes place, while at later times the temperatures for the three different cases
are very similar.

\section{Energy density fluctuations}
The main goal of this work is to study the role of the primordial QCD transition on the energy density fluctuations
in the early universe. 
If the QCD phase transition was of the first order then the speed of sound would vanish at the critical temperature, 
implying large oscillations of the energy density fluctuation. 
For the realistic case of the crossover we do expect the fluctuations to be damped because the speed of sound
never reaches zero but simply decreases to a finite value in the crossover temperature range;
nevertheless it is interesting to analyse how these fluctuations develop in presence of the crossover \cite{Schmid:1996qd}.
In order to accomplish this goal we solve the following system of equations~\cite{Schmid:1998mx}:
\begin{equation}
	\frac{1}{H}\delta'+3(c_{s}^{2}-w)\delta=\frac{k}{H}\psi-3(1+w)\alpha,
	\label{eq:S1}
\end{equation}
\begin{equation}
	\frac{1}{H}\psi'+(1-3w)\psi=-c_{s}^{2}\frac{k}{H}\delta-(1+w)\frac{k}{H}\alpha,
\end{equation}
\begin{equation}	
	\frac{1}{H}\delta'_{ew}=\frac{k}{H}\psi_{ew}-4\alpha,
\end{equation}
\begin{equation}
	\frac{1}{H}\psi'_{ew}=-\frac{k}{3H}\delta_{ew}-\frac{4k}{3H}\alpha,
\end{equation}
\begin{equation}	
\left[\left(\frac{k}{H}\right)^{2}+\frac{9}{2}(1+w_{R})\right]\alpha=-\frac{3}{2}(1+3c_{sR}^{2})\delta_{R},
\label{eq:S2}
\end{equation}
where the prime denotes the derivative with respect to the conformal time $\eta$. 
In the above equations the most important quantity for our investivation is the function $\delta\equiv\delta\varepsilon/\varepsilon$,
where $\delta\varepsilon$ correspond to the energy density fluctuation and $\varepsilon$ to the background energy density.
Among the other variables involved in the dynamics, $\psi$ is related to the fluid velocities and $\alpha$ to the
fluctuation  of the temporal part of the metric tensor. 
The function $w$ is an input and corresponds the ratio background pressure over background energy density,
which we compute case by case according to the input equation of state presented before.  
In the system of Eqs.~\eqref{eq:S1}-\eqref{eq:S2} the first two equations describe the dynamics of the QCD plasma
and follow from the energy-momentum conservation; the second couple of equations refer to the electro-weak sector
and follow from the Euler equation of general relativity; 
finally the the last one follows from the Einstein $R_{0}^{0}$-equation and couples the QCD plasma to the electroweak one.
For more details we refer to~\cite{Schmid:1996qd,Florkowski:2010mc}. 

\begin{figure}
  \includegraphics[width=7cm]{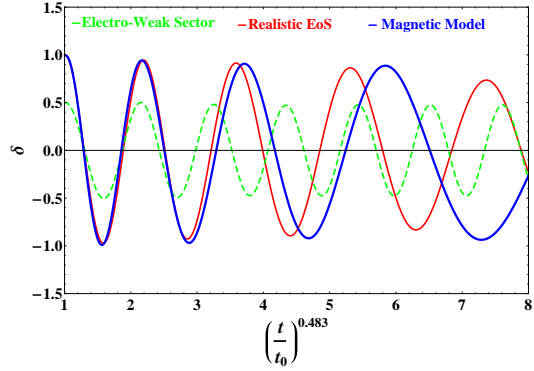}
  \caption{Time evolution of the energy density function with respect to the electro-weak part (green dashed line). The red line refers to the lattice QCD data while the blue line indicates the case in which a primordial magnetic field is included.}
  \label{fluct}
\end{figure}
In Fig.~\ref{fluct} we plot the result of our numerical solution of the system of Eqs.~\eqref{eq:S1}-\eqref{eq:S2},
focusing in particular on the energy density fluctuations $\delta$.
The initial time is $t_{0}=1.35$ $\mu$s and is chosen in order to have an initial temperature of 500 MeV, 
while the final time corresponds to 110 $\mu$s which is well beyond the QCD transition. 
In the figure the green dashed line corresponds to the electro-weak component,
while red (blue) solid line corresponds  to the QCD EoS without (with) magnetic field.
It is evident that the presence of the QCD crossover in the EoS at $B=0$ damps the energy density fluctuations
in the considered time range in both cases, affecting also the frequency of these fluctuations. 
Moreover, effect of the magnetic field is to make this damping less efficient; 
this was expected because $B\neq 0$ makes the crossover a stiffer one, thus  reducing the speed of sound 
in the crossover region compared to the speed of sound at $B=0$, and favouring the energy density fluctuations
bringing the system behaviour closer to the one expected for a first order phase transition. 

\section{Conclusions}
In conclusion, in this work we studied the effect of an equation of state which takes into account the existence
of the QCD crossover on the evolution of the early universe. In particular we computed the time evolution 
of temperature and of energy density fluctuations by using as input the lattice QCD equation of state,
describing the primordial QCD plasma at zero external fields, and a model equation of state which permits
to describe the QCD plasma in presence of a strong magnetic background.

We found that the oscillations of the energy density fluctuations for the two aforementioned cases are damped 
during the QCD crossover, this implies that there are small chances of inhomogeneities phenomena during the Big Bang nucleosynthesis.
When a strong magnetic background is added the damping is less effective,
because in the latter case the QCD crossover is stiffer and makes the confinement closer to a first order
phase transition, with a smaller speed of sound in the critical temperature range.
As an extension of this work it will be interesting to use a lattice equation of state with magnetic field 
to study our problem, like the one computed recently in \cite{Bonati:2013vba}.
Studies along this line will be reported elsewhere.


\begin{theacknowledgments}
G.~L.~G. and V.~G. acknowledge the ERC-STG funding under the QGPDyn grant.
\end{theacknowledgments}


\bibliography{proceeding}{}
\bibliographystyle{plain}


\end{document}